\documentclass[numreferences]{kluwer}

\usepackage[dvips]{epsfig}

\begin{document}

\begin{opening}

\title{\bf {Combinatorial coherent states via \\ normal ordering of bosons}}

\author{Pawel \surname{Blasiak}$^{\dag\ddag}$}
\author{Karol A. \surname{Penson}$^{\dag}$}
\author{ Allan I. \surname{Solomon}$^{\dag\S}$}
\institute{\vspace{10pt}$^{\dag}$Universit\'{e} Pierre et Marie Curie\\
Laboratoire   de  Physique   Th\'{e}orique  des  Liquides, CNRS UMR 7600\\
Tour 16, $5^{i\grave{e}me}$ \'{e}tage, 4, place Jussieu, F 75252
Paris, Cedex 05, France\\
e-mail: penson@lptl.jussieu.fr\\\vspace{5pt}
$^{\ddag}$H. Niewodnicza{\'n}ski Institute of Nuclear Physics, Polish
Academy of Science\\
Department of Theoretical Physics\\
ul. Radzikowskiego 152, PL 31-342 Krak{\'o}w, Poland
\\e-mail: Pawel.Blasiak@ifj.edu.pl\\\vspace{5pt}
$^\S$The Open University\\
 Physics and Astronomy Department\\
Milton Keynes MK7 6AA\\
e-mail: A.I.Solomon@open.ac.uk}

\begin{abstract}
We construct and analyze a family of coherent states built on
sequences of integers originating from the solution of the boson
normal ordering problem. These sequences generalize the conventional
combinatorial Bell numbers and are shown to be moments of positive
functions. Consequently, the resulting coherent states automatically satisfy 
the resolution of unity condition. In addition they display such
non-classical fluctuation properties as super-Poissonian statistics
and squeezing.
\end{abstract}

\runningtitle{Combinatorial coherent states via normal ordering of bosons}

\runningauthor{Blasiak P, Penson K A and Solomon A I}

\classification{Mathematical Subject Classifications (2000)}{81R30, 11B73.}


\end{opening}

Since their introduction in quantum optics many generalizations of
standard coherent states (CS) have been proposed.

The main purpose of such generalizations is to account for a full
description of interacting quantum systems. The
conventional CS provide a correct description of a typical
non-interacting system, the harmonic oscillator. One formal
approach to this problem is to redefine the standard boson creation
$a^\dag$ and annihilation $a$ operators, satisfying $[a,a^\dag]=1$, to
$A=af(a^\dag a)$, where the function $f(n)$, $n=a^\dag a$, is chosen to
adequately describe the interacting problem. Any deviation of $f(x)$
from $f(x)=const$ describes a non-linearity in the system. This amounts to introducing
the modified (deformed) commutation relations \cite{Solomon, Manko1, Manko2}
\begin{eqnarray}\label{0}
[A,A^\dag]=[n+1]-[n],
\end{eqnarray}
where the ``box'' function $[n]$ is defined as $[n]=nf^2(n)>0$.
Such a way of generalizing the boson commutator naturally leads to
generalized ''nonlinear'' CS in the form ($[n]!=[0][1]\ldots[n]$, $[0]=1$)
\begin{eqnarray}\label{1}
|z\rangle = \mathcal{N}^{-1/2}(|z|^2)\sum_{n=0}^\infty \frac{z^n}{\sqrt{[n]!}}|n\rangle,
\end{eqnarray}
which are eigenstates of the ``deformed'' boson annihilation operator
$A$
\begin{eqnarray}
A|z\rangle =z|z\rangle .
\end{eqnarray}
The $|n\rangle$'s are normalized kets in a Fock space.

It is worth pointing out that such nonlinear CS have been successfully
applied to a large class of physical problems in quantum optics
\cite{Vogel1, Vogel2}. The comprehensive treatment of CS of
the form of Eq. (\ref{1}) can be found in \cite{Solomon, Manko1, Manko2}. 

An essential ingredient in the definition of CS is the completeness property (or the resolution of unity
condition) \cite{Klauder, KPS1, KPS2}.
A guideline for the construction of CS in general has been put forward in \cite{Klauder} as a minimal set of conditions. Apart from the conditions of normalizability and
continuity in the complex label $z$, this set reduces to satisfaction of the  resolution of unity condition. This implies the existence of a positive function $\tilde{W}(|z|^2)$ satisfying
\begin{eqnarray}\label{A}
{\int\int}_{\mathbb{C}}d^2z\ |z\rangle \tilde{W}(|z|^2)\langle z|=I=\sum_{n=0}^\infty|n\rangle \langle n|,
\end{eqnarray}
which reflects the completeness of the set $\{|z\rangle\}$.

In Eq.(\ref{A}) $I$ is the unit operator and $|n\rangle$ is a complete set of
orthonormal eigenvectors. In a general
approach one chooses strictly positive parameters $\rho (n),\ n=0,1,\ldots$
such that the state $|z\rangle$ which is normalized, $\langle z|z\rangle =1$, is given by
\begin{eqnarray}\label{I}
|z\rangle =\mathcal{N}^{-1/2}(|z|^2)\sum_{n=0}^\infty \frac{z^n}{\sqrt{\rho(n)}}|n\rangle,
\end{eqnarray}  
with normalization
\begin{eqnarray}\label{J}
\mathcal{N}(|z|^2)=\sum_{n=0}^\infty \frac{|z|^{2n}}{\rho(n)}>0,
\end{eqnarray}
which in this note we assume to be a convergent series in $|z|^2$ for all $z\in\mathbb{C}$.
In view of Eqs. (\ref{0}) and (\ref{1}) this corresponds to
$\rho(n)=[n]!$ or $[n]=\rho(n)/\rho(n-1)$ for $n=1,2,\ldots$.

Condition (\ref{A}) can be shown to be equivalent to the following
infinite set of equations \cite{KPS1}:
\begin{eqnarray}\label{B}
\int_0^\infty x^n\left[\pi \frac{\tilde{W}(x)}{\mathcal{N}(x)}\right]dx=\rho(n),\ \ n=0,1,\ldots,
\end{eqnarray}
a Stieltjes moment problem for
$W(x)=\pi\frac{\tilde{W}(x)}{\mathcal{N}(x)}$.

Recently considerable progress was made  in finding
explicit solutions of Eq.(\ref{B}) for a large set of $\rho(n)$'s, generalizing the
conventional choice $|z\rangle_c$ for which  $\rho_c(n)=n!$ with $\mathcal{N}_c(x)=e^x$ (see
Refs. \cite{KPS1, KPS2, Quesne, Vourdas, Odzijewicz, Gazeau} and
references therein), thereby extending the known families of CS. This
progres was faciliated by the observation that when the moments form
certain combinatorial sequences a solution of the associated Stieltjes
moment problem may be obtained explicitly \cite{PS}.

In this work we make contact with the combinatorial sequences appearing in
the solution of the boson normal ordering problem \cite{BPS1, AoC}. These sequences
have the very desirable property of being moments of Stieltjes-type
measures and so automatically fulfill the resolution of unity
requirement. It is therefore natural to use these sequences for the
CS construction, thereby providing a link between the quantum states
and the combinatorial structures. 

The normal ordering problem for canonical bosons  $[a,a^\dag]=1$ is
related to certain combinatorial numbers $S(n,k)$ called Stirling
numbers of the second kind through \cite{Katriel}
\begin{eqnarray}
(a^\dag a)^n=\sum_{k=1}^nS(n,k) (a^\dag)^k a^k,
\end{eqnarray}
with corresponding numbers $B(n)=\sum_{k=1}^n S(n,k)$ called Bell numbers.

For integers
$n,r,s>0$ we define generalized Stirling numbers of the second
kind $S_{r,s}(n,k)$ through ($r\geq s$):
\begin{eqnarray}\label{C}
[(a^\dag)^ra^s]^n=(a^\dag)^{n(r-s)}\sum_{k=s}^{ns}S_{r,s}(n,k)(a^\dag)^ka^k,
\end{eqnarray}
as well as generalized Bell numbers $B_{r,s}(n)$ 
\begin{eqnarray}\label{D}
B_{r,s}(n)=\sum_{k=s}^{ns}S_{r,s}(n,k).
\end{eqnarray}
For both $S_{r,s}(n,k)$ and $B_{r,s}(n)$ exact and explicit formulas
have been found \cite{BPS1, AoC}.

In this study we shall only be interested in the subset  $B_{r,1}(n)$,
for which a convenient infinite series representation may be given
\cite{BPS1, AoC}:
\begin{eqnarray}\label{E}
B_{r,1}(n)=\frac{(r-1)^{n-1}}{e}\sum_{k=0}^\infty 
\frac{1}{k!}\frac{\Gamma(n+\frac{k+1}{r-1})}{\Gamma(1+\frac{k+1}{r-1})},\ \ r>1.
\end{eqnarray}
This is a generalization of the celebrated Dobi\'nski relation 
\begin{eqnarray}\label{M}
B_{1,1}(n)=\frac{1}{e}\sum_{k=0}^\infty\frac{k^n}{k!}
\end{eqnarray}  
for conventional Bell numbers $B_{1,1}(n)$, see \cite{Wilf},\cite{Pitman}. For reference
we quote $B_{1,1}(n)=1,2,5,15,52,203,\ldots$.

\begin{figure}[t]
\vspace{1cm}
\begin{center}\resizebox{10cm}{!}{\includegraphics{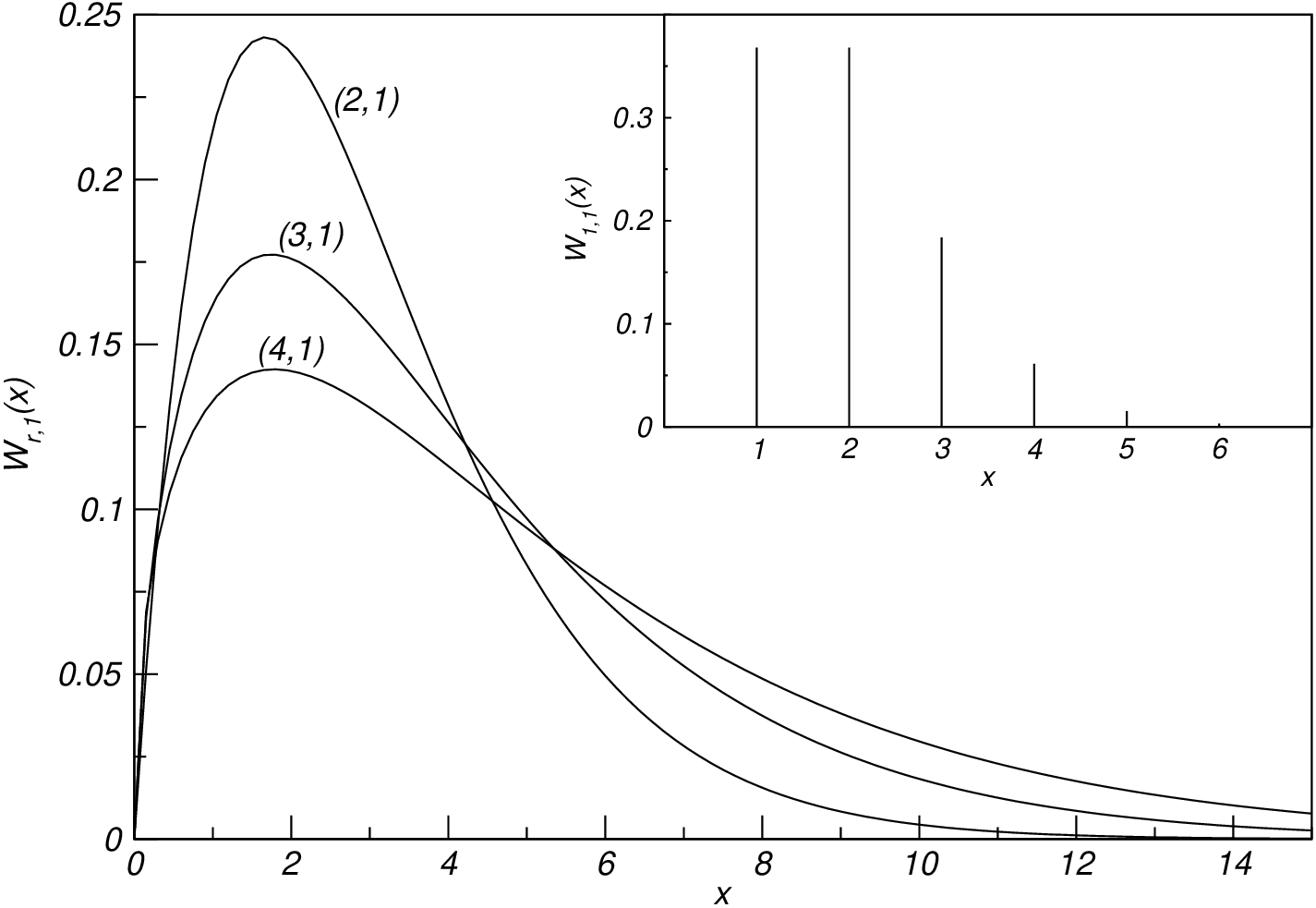}}
\caption{The weight functions $W_{r,1}(x)$, ($x=|z|^2$), in
the resolution of unity for $r=2,3,4$ (continuous
curves) and for $r=1$, a  Dirac's comb (in the inset), as a function of $x$.}\label{FigA}
\end{center}
\end{figure}

From Eqs.(\ref{C}) and (\ref{D}) one sees that $B_{r,1}(n)$ are integers. Eq.(\ref{E}) gives explicitly $B_{2,1}(n)=1,3,13,73,501,4051\ldots$; $B_{3,1}(n)=1,4,25,211,2236,28471,\ldots$; $B_{4,1}(n)=1,5,41,465,6721,117941\ldots$ etc. 
Through transformations of
Eq.(\ref{E}) one finds that $B_{r,1}(n)$ for $r>1$ can be expressed as special
values of hypergeometric functions of type $_1F_{r-1}$:
\begin{eqnarray}
&&B_{2,1}(n)=\frac{n!}{e}\ {_1F_1}(n+1;2;1),\\
&&B_{3,1}(n)=\frac{2^{n-1}}{e}\left(\frac{2\Gamma(n+\frac{1}{2})}{\sqrt{\pi}}\
{_1F_2}(n+\frac{1}{2};\frac{1}{2},\frac{3}{2};\frac{1}{4})\right.\nonumber\\
&&\ \ \ \ \ \ \ \ \ \ \ \ \ \ \ \ \ \ \ \ \ \ \ \ \left.+n!\ {_1F_2}(n+1;\frac{3}{2},2;\frac{1}{4})\right),\\
&&etc.\nonumber
\end{eqnarray}  

It is essential for our purposes to observe that the integer
$B_{r,1}(n+1)$, $n=0,1,\ldots$ is the $n$-th moment of a positive function
$W_{r,1}(x)$ on the positive half-axis. For $r=1$ we see from
Eq.(\ref{M}) that $B_{1,1}(n+1)$ is the $n$-th moment of a discrete
distribution $W_{1,1}(x)$ located at positive integers, a so-called
{\em Dirac comb}:
\begin{eqnarray}\label{Y}
B_{1,1}(n+1)=\int_0^\infty x^n\left[\frac{1}{e}\sum_{k=1}^\infty
\frac{\delta(x-k)}{(k-1)!}\right]\ dx.
\end{eqnarray} 
For every $r>1$ a continuous distribution $W_{r,1}(x)$ will be
obtained by excising $(r-1)^n\ \Gamma(n+\frac{k+1}{r-1})$ from Eq.(\ref{E}),
performing the inverse Mellin transform on it and inserting the result
back in the sum of Eq.(\ref{E}). (Note also that
$B_{r,1}(0)=\frac{e-1}{e}$, $r=2,3,\ldots$ is no longer
integral). In this way we obtain
\begin{eqnarray}\label{N}
B_{r,1}(n+1)=\int_0^\infty x^nW_{r,1}(x)\ dx,
\end{eqnarray}
which yields for $r=2,3,4$:

\begin{eqnarray}
&&W_{2,1}(x)=e^{-x-1}\sqrt{x}\ I_1(2\sqrt{x}),\label{F}\\
&&W_{3,1}(x)=\frac{1}{2}\sqrt{\frac{x}{2}}e^{-\frac{x}{2}-1}\left(\frac{2}{\sqrt{\pi}}\
{_0F_2}(\frac{1}{2},\frac{3}{2};\frac{x}{8})+\frac{x}{\sqrt{2}}\ {_0F_2}(\frac{3}{2},2;\frac{x}{8})\right),\label{G}\\
&&W_{4,1}(x)=\frac{1}{18\pi\Gamma(\frac{2}{3})}e^{-\frac{x}{2}-1}\left(3^{\frac{13}{6}}\Gamma^2(\frac{2}{3})x^{\frac{1}{3}}\
{_0F_3}(\frac{1}{3},\frac{2}{3},\frac{4}{3};\frac{x}{81})+\right.\\
&&\ \ \ \ \ \ \ \ \ \ \ \ \ \ \left.3^{\frac{4}{3}}\pi
x^{\frac{2}{3}}\
{_0F_3}(\frac{2}{3},\frac{4}{3},\frac{5}{3};\frac{x}{81})+\pi\Gamma(\frac{2}{3})x\
{_0F_3}(\frac{4}{3},\frac{5}{3},2;\frac{x}{81})\right)\label{H}.\nonumber
\end{eqnarray}
In Eqs.(\ref{F}),(\ref{G}),(\ref{H}) $I_\nu (y)$ and $_0F_p(\ldots;y)$ are modified Bessel and hypergeometric functions, respectively. 
Other $W_{r,1}(x)>0$ for $r>4$ can be generated by essentially the same
procedure.

In Figure \ref{FigA} we display the weight functions $W_{r,1}(x)$ for
$r=1\ldots4$; all of them are normalized to one. In the inset the
height of the vertical line at $x=k$ symbolizes the strength of the
delta function $\delta(x-k)$, see Eq.(\ref{Y}). For further properties of
$W_{1,1}(x)$ and more generally of $W_{r,r}(x)$ associated with
Eq.(\ref{D}), see \cite{BPS2}. 

A comparison of Eqs.(\ref{I}),(\ref{J}) and (\ref{N}) indicates that the
normalized states defined through $\rho(n)=B_{r,1}(n+1)$ as
\begin{eqnarray}\label{Z}
|z\rangle_r =\mathcal{N}_r^{-1/2}(|z|^2)\sum_{n=0}^\infty \frac{z^n}{\sqrt{B_{r,1}(n+1)}}|n\rangle,
\end{eqnarray}  
with normalization
\begin{eqnarray}
\mathcal{N}_r(x)=\sum_{n=0}^\infty \frac{x^n}{B_{r,1}(n+1)}>0,
\end{eqnarray}
{\it automatically} satisfy the resolution of unity
condition of Eq.(\ref{A}), since for
$W_{r,1}(x)=\pi\frac{\tilde{W}_{r,1}(x)}{\mathcal{N}_r(x)}$:
\begin{eqnarray}
{\int\int}_{\mathbb{C}}d^2z\ |z\rangle_r \tilde{W}_{r,1}(|z|^2){_r\langle z|}=I=\sum_{n=0}^\infty|n\rangle \langle n|.
\end{eqnarray}
Note that Eq.(\ref{Z}) is equivalent to Eq.(\ref{1}) with the
definition $[n]_r=B_{r,1}(n+1)/B_{r,1}(n)$, $n=0,1,2,\ldots$.

\begin{figure}[t]
\vspace{1cm}
\begin{center}\resizebox{10cm}{!}{\includegraphics{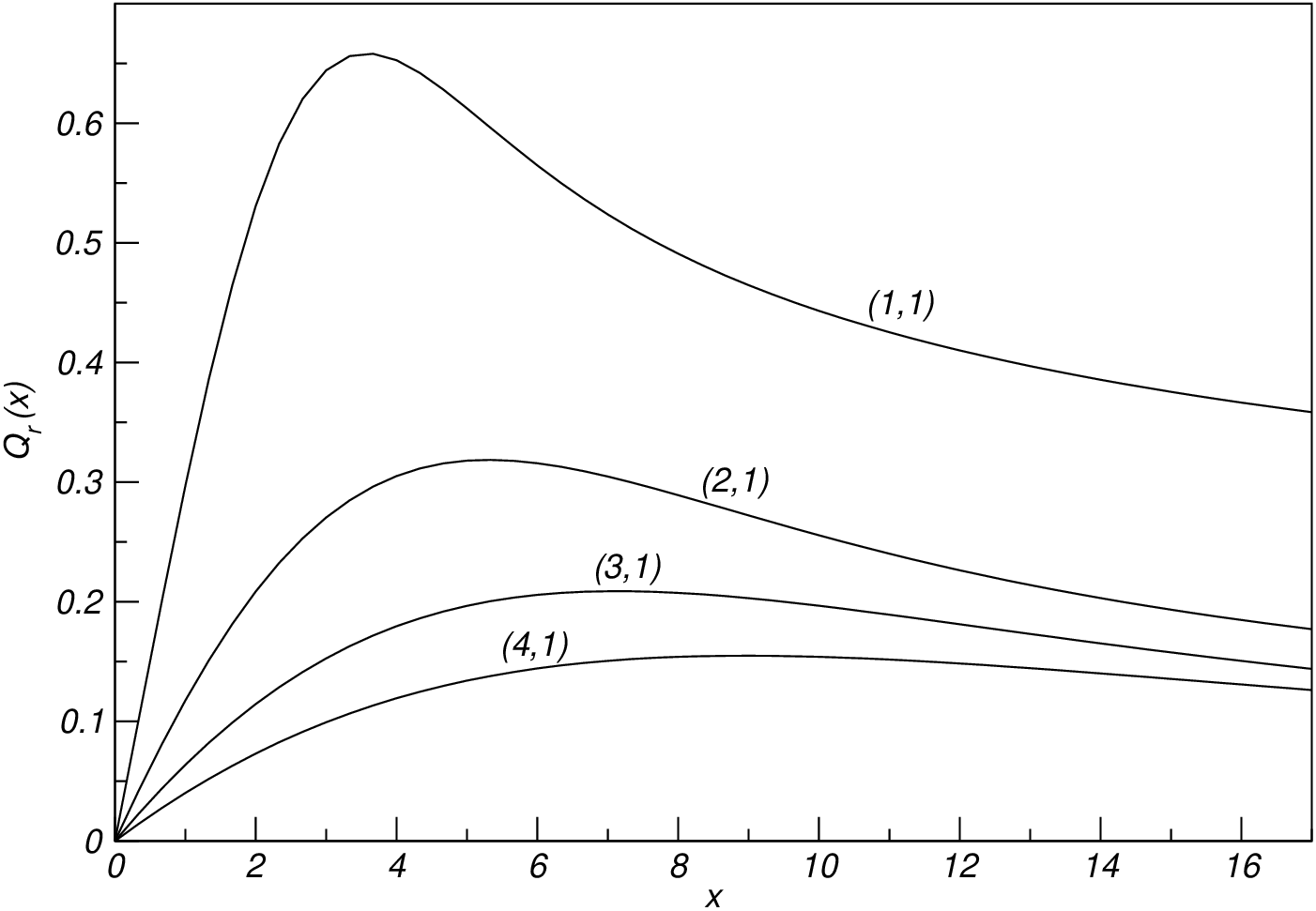}}
\caption{Mandel parameters $Q_r(x)$, for
$r=1\ldots4$, as a function of $x=|z|^2$, see Eq.(\ref{Q}).}\label{FigB}
\end{center}
\end{figure}

Having satisfied the completeness condition with the functions
$W_{r,1}(x)$, $r=1,2,\ldots$ we now proceed to examine the quantum-optical
fluctuation properties of the states $|z\rangle_r$.
From now on we consider the $|n\rangle$'s to be eigenfunctions of the boson
number operator $N=a^\dag a$, i.e. $N|n\rangle=n|n\rangle$.
The Mandel parameter \cite{KPS1}
\begin{eqnarray}\label{Q}
Q_r(x)=x\left(\frac{{\mathcal{N}_r}^{\prime\prime} (x)}{{\mathcal{N}_r}^\prime
(x)}-\frac{{\mathcal{N}_r}^\prime (x)}{\mathcal{N}_r(x)}\right),
\end{eqnarray}
allows one to distinguish between the sub-Poissonian (antibunching
effect, $Q_r<0$) and
super-Poissonian (bunching effect, $Q_r>0$) statistics of the beam. In
Figure \ref{FigB} we display $Q_r(x)$ for $r=1\ldots4$.
It can be seen that all the states $|z\rangle_r$ in question are
super-Poissonian in nature, with the deviation from $Q_r=0$,
which characterises the conventional CS, diminishing for $r$ increasing.
\begin{figure}[t]
\vspace{1cm}
\begin{center}\resizebox{10cm}{!}{\includegraphics{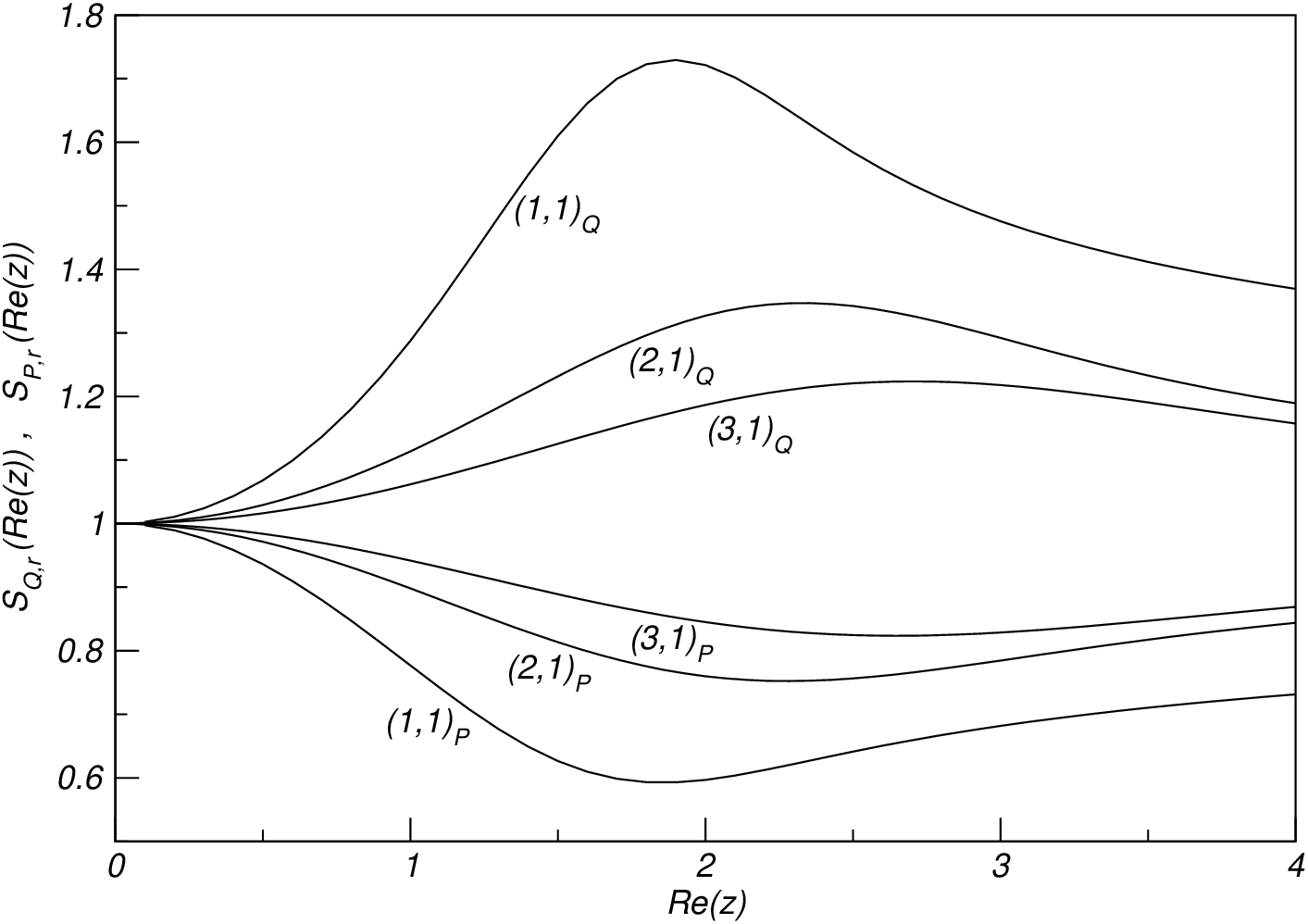}}
\caption{The squeezing parameters of Eqs.(\ref{SQ}) and (\ref{SP}) for the
coordinate $Q$ (three upper curves) and for the momentum $P$ (three
lower curves) for different $r$, as a function of $Re(z)$, for $r=1,2,3$.}\label{FigC}
\end{center}
\end{figure}

In Figure \ref{FigC} we show the behavior of 
\begin{eqnarray}\label{SQ}
S_{Q,r}(z)=\frac{{_r\langle z|(\Delta Q)^2|z\rangle_r}}{2},
\end{eqnarray}
and 
\begin{eqnarray}\label{SP}
S_{P,r}(z)=\frac{{_r\langle z|(\Delta P)^2|z\rangle_r}}{2},
\end{eqnarray}
which are the measures of squeezing in the coordinate and  momentum quadratures
respectively. In the display we have chosen the section along $Re(z)$. All the states $|z\rangle_r$ are squeezed in the momentum $P$ and
dilated in the coordinate $Q$. The degree of squeezing and dilation diminishes
with increasing $r$. By introducing the imaginary part in $z$ the
curves of $S_Q(z)$ and $S_P(z)$ smoothly transform into one another,
with the identification $S_Q(i\alpha)=S_P(\alpha)$ and $S_P(i\alpha)=S_Q(\alpha)$ for any
positive $\alpha$.

\begin{figure}[p]
\vspace{1cm}
\begin{center}\resizebox{10cm}{!}{\includegraphics{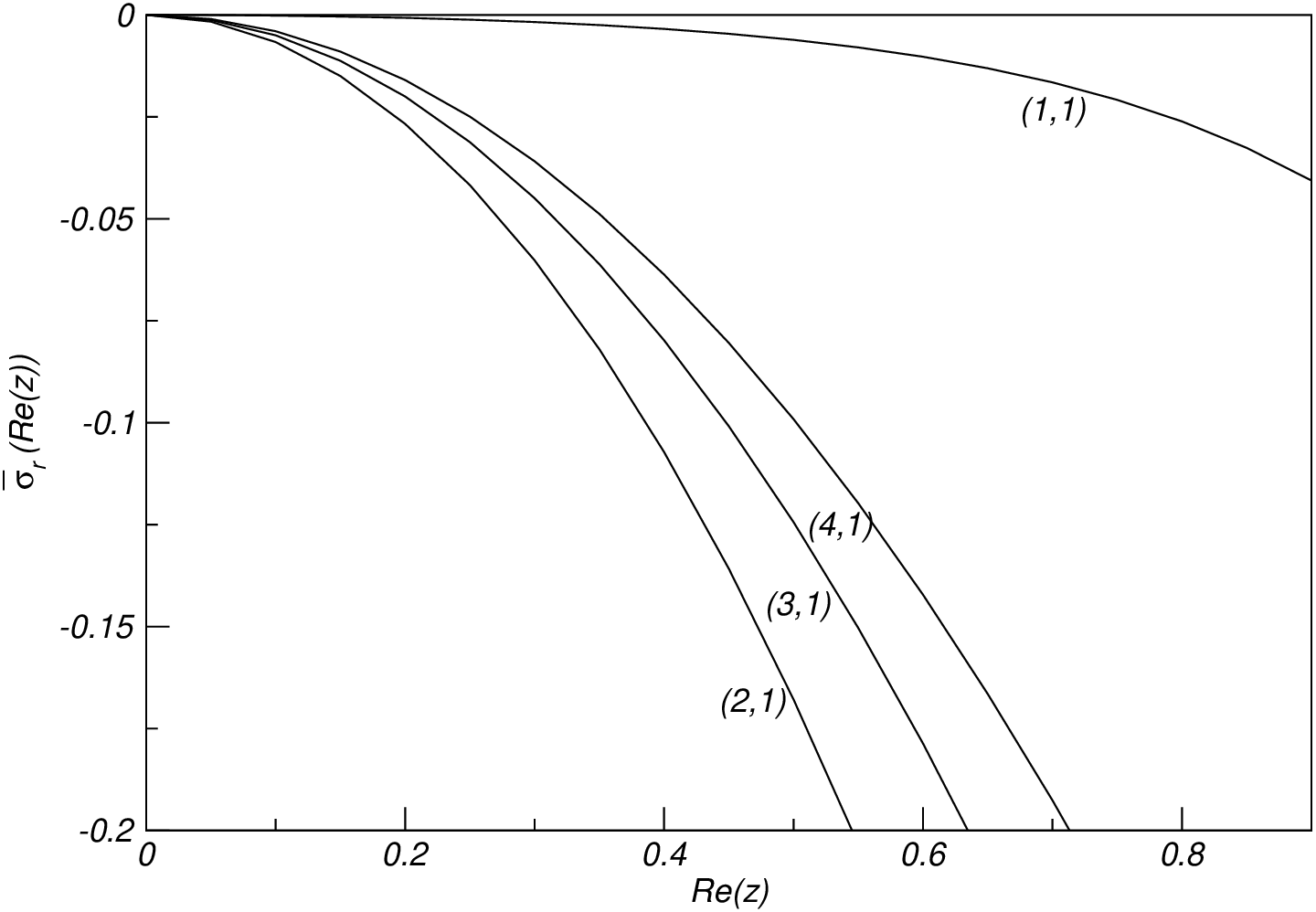}}
\caption{The signal-to-quantum noise ratio relative
to its value in the standard coherent states $\bar{\sigma}_r$, see Eq.(\ref{O}), as a
function of $Re(z)$, for $r=1\ldots4$.\bigskip}\label{FigD}
\end{center}
\end{figure}

\begin{figure}
\begin{center}

\resizebox{10cm}{!}{\includegraphics{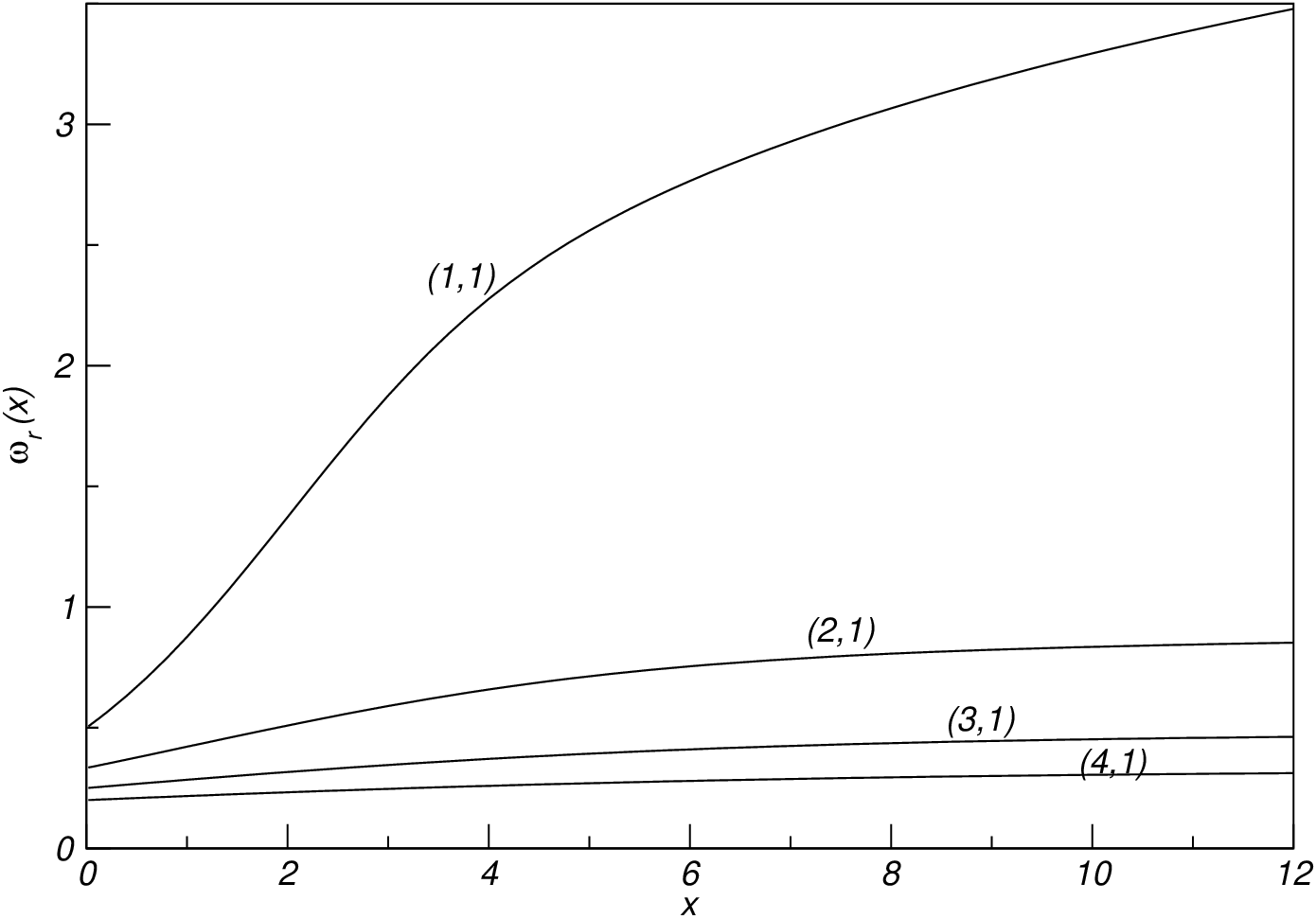}}
\caption{Metric factor $\omega_r(x)$, calculated with
Eq.(\ref{W}) as a function of $x=|z|^2$, for $r=1\ldots4$.}\label{FigE}
\end{center}
\end{figure}
In Figure \ref{FigD} we show the signal-to-quantum noise ratio \cite{Yuen} relative
to $4[{_c\langle z|N|z\rangle_c}]=4|z|^2$, its value in conventional coherent states;
i.e the quantity $\bar{\sigma}_r=\sigma_r-4[{_c\langle z|N|z\rangle_c}]$, where
\begin{eqnarray}\label{O}
\sigma_r=\frac{[{_r\langle z|Q|z\rangle_r}]^2}{(\Delta Q)^2},
\end{eqnarray}
with $(\Delta Q)^2={_r\langle z|Q^2|r\rangle_r}-[{_r\langle z|Q|r\rangle_r}]^2$. Again only
the section $Re(z)$ is shown. We conclude from Figure \ref{FigD} that the states $|z\rangle_r$ are more ``noisy'' than the standard CS with $\rho_c(n)=n!$ .

In Figure \ref{FigE} we give the metric factors
\begin{eqnarray}\label{W}
\omega_r(x)=\left[x\frac{{\mathcal{N}_r}^\prime (x)}{\mathcal{N}_r(x)}\right]^\prime ,
\end{eqnarray}
which describe the geometrical properties of embedding the surface of
coherent states in Hilbert space, or equivalently a measure of a
distortion of the complex plane induced by the CS \cite{KPS1}. Here,
as far as $r$ is concerned, the state $|z\rangle_1$ appears to be most
distant from the $|z\rangle_c$ CS for which $\omega_c=1$.

\begin{figure}[p]
\vspace{1cm}
\begin{center}\resizebox{10cm}{!}{\includegraphics{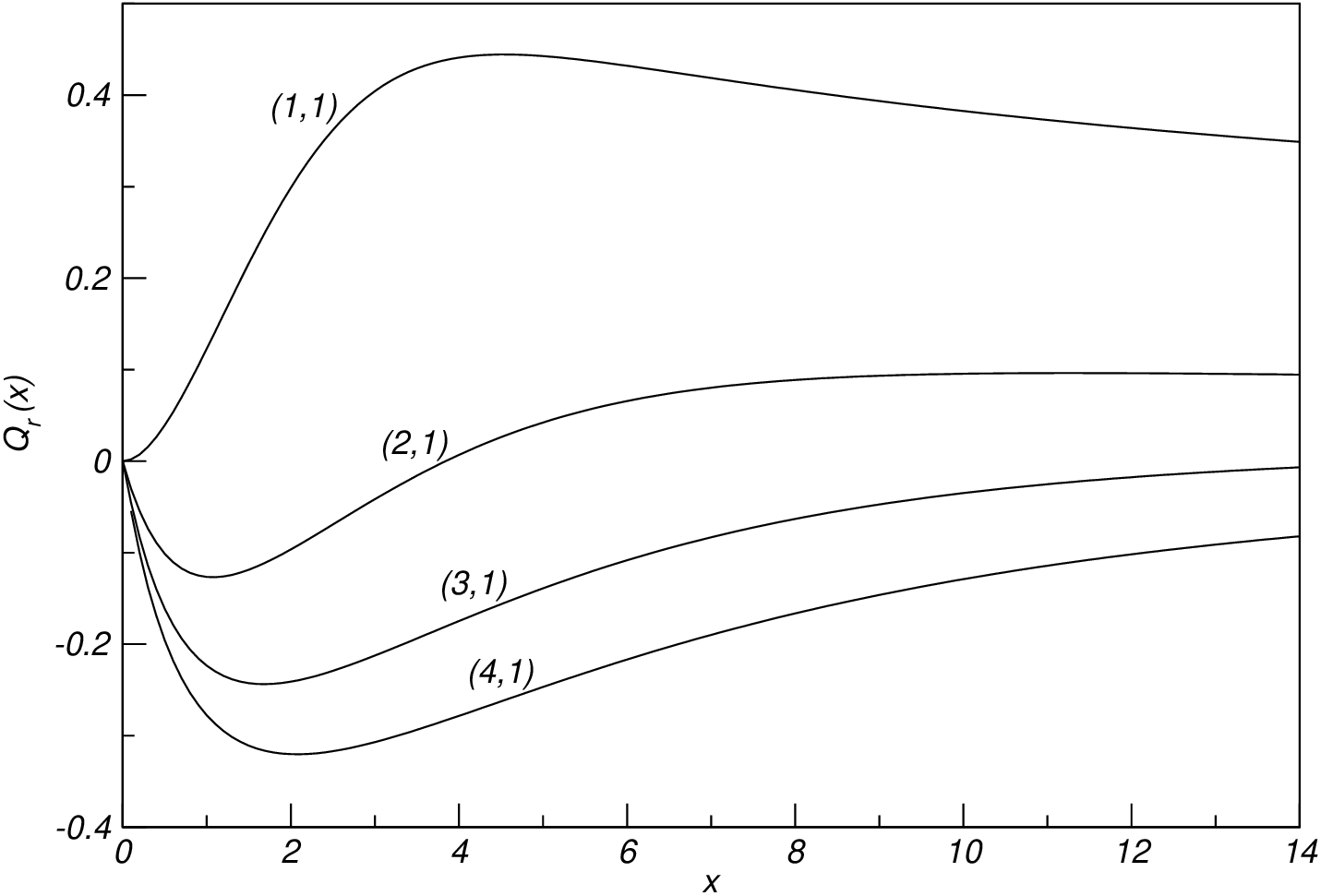}}
\caption{Mandel parameters $Q_r(x)$ for
$r=1\ldots4$, as a function of $x=|z|^2$, for states with
$\rho(n)=B_{r,1}(n)$.\bigskip}\label{FigF}
\end{center}
\end{figure}

\begin{figure}
\begin{center}

\resizebox{10cm}{!}{\includegraphics{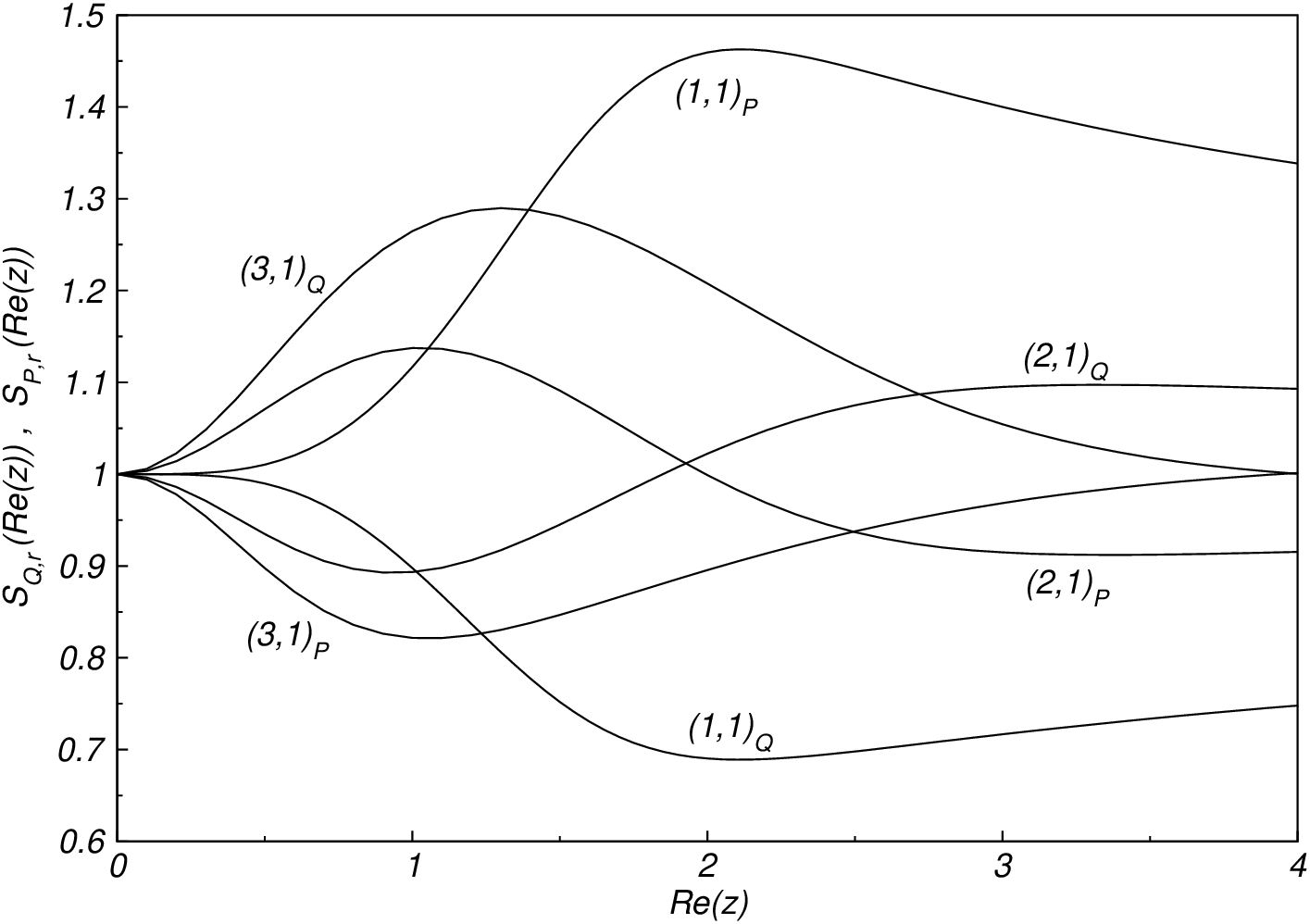}}
\caption{Squeezing parameters for the states with
$\rho(n)=B_{r,1}(n)$ for $r=1,2,3$, as a function of $x=|z|^2$. The
subscripts $P$ and $Q$ refer to momentum and coordinate variables respectively.}\label{FigG}
\end{center}
\end{figure}

The use of sequences $B_{r,1}(n)$ to construct CS is not limited
to the case exemplified by Eq.(\ref{Z}). In fact, any sequence
of the form $B_{r,1}(n+p),\ p=0,1,\ldots$ will also define a set of CS, as
then their respective weight functions will be
$V_{r,1}^{(p)}(x)=x^{p-1}W_{r,1}(x)>0$. Will the physical properties
of CS defined with $\rho_p(n)=B_{r,1}(n+p)$  depend sensitively on $p$ ?
Our first guess was that by varying $p$ the salient features of the
physical results will not change. It was based on a study of
$\rho(n)=(n+p)!$ in \cite{KPS1} as a function of $p$ in a somewhat
similar situation. However, this was not confirmed by actual
calculations (and in fact could have been envisaged from the outset as
$V^{(0)}_{r,1}(x)$ looks quite different from $W_{r,1}(x)$). A case in point is that of $p=0$ for which qualitative differences from Figure \ref{FigB} ($p=1$) appear. In Figure \ref{FigF} we present the Mandel parameter for these
states. Whereas for $r=1$ the state is still super-Poissonian, for
$r=2,3,4$ one observes novel behaviour, namely a crossover from sub-
to super-Poissonian statistics for finite values of $x$. Also,
as indicated in Figure \ref{FigG}, we note a cross-over between
squeezing and dilating behaviour for different $r$. These curious
features merit further investigation.

In conclusion, we have used the combinatorial numbers $B_{r,1}(n)$ 
to construct and analyze families of coherent states which automatially  satisfy the
resolution of unity property. This study will be extended to other $B_{r,s}(n)$
arising in the problem of boson normal ordering, since such sequences,
being moments of positive functions \cite{BPS1, AoC}, are ideally suited for the construction of coherent states.


\begin{acknowledgements}

We thank A. Horzela, M. Mendez, J. Pitman, C. Quesne and A. Vourdas for important discussions.

\end{acknowledgements}

\end{document}